Ulrich Menne[1]
Christian Raack[*]
Roland Wessäly[1]
Daniel Kharitonov[2]

# Optimal Degree of Optical Circuit Switching in IP-over-WDM Networks[**]

[1] atesio GmbH, Berlin, Germany
[2] Juniper Networks, USA

[*] This research has been supported by the DFG research Center Matheon.

# Optimal Degree of Optical Circuit Switching in IP-over-WDM Networks

**Ulrich Menne, Christian Raack, Roland Wessäly, Daniel Kharitonov**

*atesio GmbH, Zuse Institute Berlin (ZIB), Juniper Networks*

{*menne,wessaely*}*@atesio.de, raack@zib.de, dkh@juniper.net*

**Abstract:**
In this paper, we study the influence of technology, traffic properties and price trends on optimized design of a reference IP-over-WDM network with rich underlying fiber topology. In each network node, we investigate the optimal degree of traffic switching in an optical (lambda) domain versus an electrical (packet) domain, also known as measure of *node transparency*. This measure is studied in connection to changes in traffic volume, demand affinity, optical circuit speeds and equipment cost. By applying variable design constraints, we assess the relative roles of the two distinct equipment groups, IP routers and optical cross-connects, with respect to resulting changes in cost-sensitive network architectures.

**OCIS codes:** 060.4510, 060.4254, 060.4256

## 1. Introduction

The use of optical switching technology to supplement the growth of packet-processing (IP/MPLS) nodes has been widely investigated since the late 1990s [1] and found practical applications in wavelength-division multiplexing (WDM) networks with cross-connect capabilities. Whether optimized heuristically or programmatically, the multi-layer network combining packet switching in IP routers and lambda switching in the WDM layer invariably presents the problem of balance between processing transit traffic in optical and electrical domains (degree of network *transparency*). While every reference network is unique, numerous studies have shown that introduction of optical circuit switching in certain (formerly) full-mesh routing topologies can reduce network capital costs by 20 to 30 percent [2] [3] [4]. However, there is very little research on the general subject of optimal network transparency in relation to common design parameters. This gap prevents practical network designers and engineers from understanding the impact of network growth in relation to rapidly shifting technology trends.

Throughout this paper, we employ a reference fiber-rich network to study cost-optimal architectures with respect to the influence of varying technology, demand and cost patterns. In the course of optimization, we test the impact of the following variables on network cost and transparency:

  (i)  traffic volume

 (ii)  traffic pattern and distribution

(iii)  available circuit speeds

(iv)  cost for individual wavelengths

To compute feasible solutions for each design variation, we apply an extension of the original two-layer optimization algorithm [4] capable of producing optimal or near-optimal results (maximum uncertainty 5 percent). We use these results to research the "what-if" scenarios of network evolution and optimal placement of packet (IP/MPLS) versus optical switching equipment.

This paper is organized as follows: *Section 2* introduces the used reference data, that is, the considered architectures, networks and traffic models. Also the node opacity measure is defined. *Section 3* reveals our computations and sensitivity analysis together with our findings and observations. *Appendix A* provides a description of the used hardware equipment and their relations together with a detailed cost model. *Appendix B* explains our mathematical model used to obtain a cost-optimal network solution.

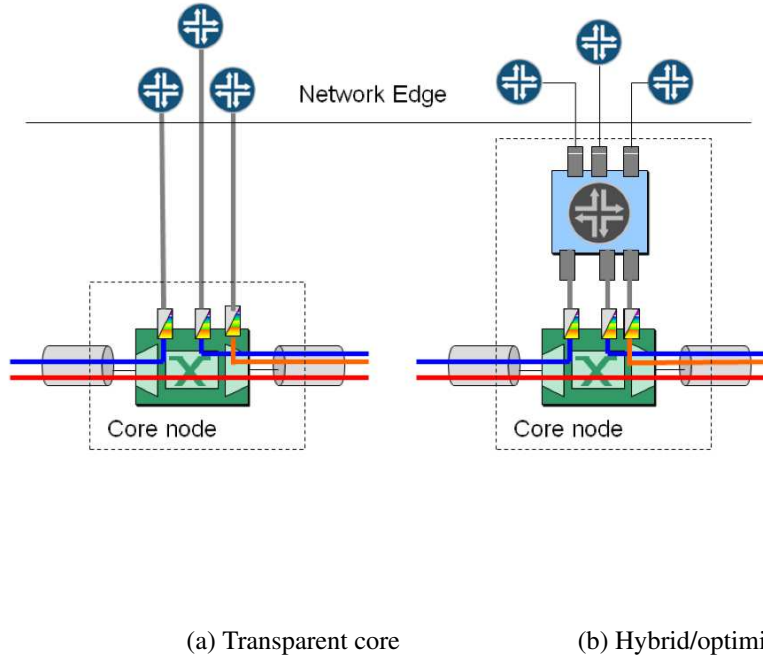

(a) Transparent core  (b) Hybrid/optimized

Fig. 1. Node architectures

## 2. Baseline Architecture and Reference Data

In our study, we consider a generic, two-layer packet-over-WDM architecture designed to support IP/MPLS payloads. We expect all traffic demands to originate from ingress edge devices and pass through intermediate network nodes toward the egress edge devices. We assume every node can support a combination of IP and WDM gear and carry local (add-drop) as well as transit traffic (in electrical or optical domain) as deemed necessary by the planning algorithm.

### 2.1. Node Architecture

In principle, we assume two different network architectures with respect to the core functions.

In the first architecture, we consider a case where the core is purely optical and all edge demands can be placed directly onto wavelengths – see Figure 1(a). In that case, the network provisions only optical circuits and there is no electrical multiplexing layer for IP/MPLS traffic (no core routers are present). We refer to this architecture as a "fully transparent" (or *transparent core*) solution.

In the second architecture, IP/MPLS demands originating on respective edge router(s) can be multiplexed by the core IP device and groomed into lightwaves toward the WDM switch – the latter also known as optical cross-connect. In that node type, both the core router and optical cross-connect can handle transit traffic, so the network (as a whole) has at least some degree of opacity – see Figure 1(b). We refer to such solutions as hybrid/optimized (or simply *optimized*) as in [2] Notice that our definition of a "transparent core" solution differs from transparency definitions given in [2] and [4]. In the transparent solutions in [2] the IP core router is always present even if it never carries transit traffic. In [4], hybrid/optimized solutions based on a node architecture similar to Figure 1(b) are called "transparent solutions".

In both architectures we allow the use of additional optical-only nodes where no IP/MPLS demands terminate. At these nodes only WDM cross-connects – without any additional IP routing functionality – can be added if needed by the optimization algorithm. In all cases, edge router(s) are not part of the core nodes. However, we do calculate the cost of core-facing interfaces on the edge devices whenever the core architecture might affect it.

### 2.2. Node Transparency and Opacity

In this work we aim at understanding the impact of network design parameters on optimal multi-layer designs. We hence introduce the reciprocal terms "transparency" and "opacity" to denote the ratio of packet versus lambda switching. To define the proper measure for this ratio, we need to ensure that network nodes in Figure 1(a) turn out to be

100% transparent (or equivalently 0% opaque). The transparency of a node as in Figure 1(b) however is supposed to be between 0% and 100% and depends on the amount of transit traffic in the IP layer versus WDM layer. In this respect we introduce the following terminology for every node location $i$ in a network solution:

- We denote by $F_{IP}(i)$ the total *IP transit traffic* at node $i$, that is, $F_{IP}(i)$ refers to all IP traffic in Gbps that is switched at the IP router at location $i$ but that is *not* terminated (as demand flows) toward the edge. Clearly, such flows are entering and leaving the IP router from/toward the core. However, we count every flow unit only once in $F_{IP}(i)$. It holds $F_{IP}(i) = 0$ if no IP router is installed.

- Similarly, we denote by $F_{WDM}(i)$ the total *WDM transit traffic* in Gbps at node $i$, that is, $F_{WDM}(i)$ refers to all IP traffic at node $i$ that is not switched toward the installed IP core router or toward the installed edge router. This traffic is hence carried on wavelengths that are optically bypassed.

With these transit flow definitions we refer to the *opacity* $\phi(i)$ of node $i$ as the following

$$\phi(i) := \frac{100 \cdot F_{IP}(i)}{F_{IP}(i) + F_{WDM}(i)} \quad (1)$$

Assuming $F_{IP}(i) + F_{WDM}(i) > 0$, a node is said to be opaque if $\phi(i) = 100\%$. In this case there are no transit flows at the optical cross-connect of node $i$. Instead, all demand flows are terminated or switched at the IP-layer. In contrast, if $\phi(i) = 0\%$ node $i$ is said to be transparent, with all transit traffic passing through the node without processing in the core router. In case $F_{IP}(i) + F_{WDM}(i) = 0$ the opacity status of $i$ is undefined. In this case there is only demand flow at $i$. That is, the node acts as a leaf node in the network with no transit traffic and all traffic is terminated toward the edge.

For the definition of the *network opacity* $\phi$ we sum up the transit traffic values over all IP points of presence (PoP) locations in the network, that is, $\phi$ is given by the following:

$$\phi := \frac{100 \cdot F_{IP}}{F_{IP} + F_{WDM}} \quad (2)$$

where $F_{IP}$ (respectively $F_{WDM}$) refers to the total IP (respectively WDM) traffic of the PoP locations. Notice that in this paper we optimize networks with respect to the cost for capacity. As the node flows depend on the actual routings, there might be different opacity values possible for the same optimal capacity solution. However, we still believe that $\phi$ provides a meaningful measure to compare the characteristics of different solutions.

*2.3. Reference Data*

Our study has been performed on a German fiber reference network with 50 WDM node locations and 89 links, see Fig. 2(b). Only 17 out of 50 nodes are PoP locations with IP/MPLS demand sources and sinks, so those nodes have edge routers and can also be equipped with IP core routers if needed. Therefore, this reference network is fiber-rich and well provisioned for optical switching.

**Device definitions and costs**  Since we chose to optimize network design based on capital cost, using a solid inventory is critical. In our work, we employ a cost model similar to that developed in the European project NOBEL [5,6] with several improvements. For the detailed description of changes refer to *Appendix A*.

**Common traffic characteristics**  In this work, we assume all traffic sources and sinks to be static and producing symmetric, constant bit-rate streams. Such a model allows for easy demand parametrization and practically corresponds to network sizing based on peak demands. The distribution of demands between destinations forms a traffic matrix, where every node acts as both source and sink.

**Input design parameter - traffic gravity**  Traffic requirements are given in the IP layer between the 17 PoP-locations of the reference network – see Fig. 2(b). For our evaluations we use a *centralized* (DFN) and a *decentralized* (DWG) point-to-point traffic matrix, based on traffic measurements and population statistics, respectively.

- The centralized DFN matrix is calculated from real-life accounting in the years 2004 and 2005 in the German research network operated by the German DFN-Verein [7]. The corresponding data sets have been uploaded recently to the SNDlib (Survivable Network Design Library) [8]. We used a data set that states the monthly byte

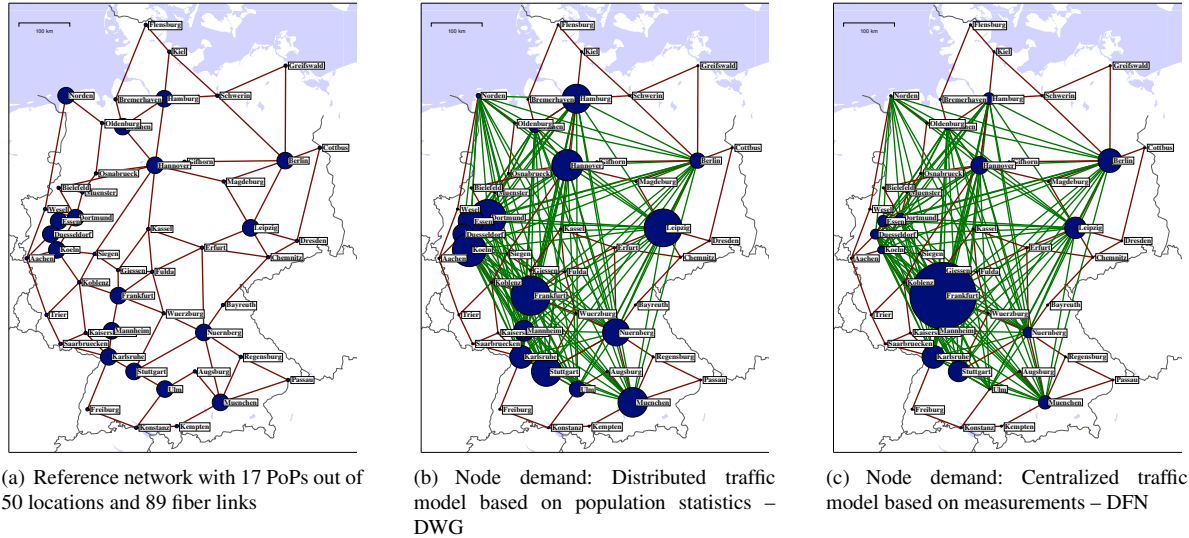

(a) Reference network with 17 PoPs out of 50 locations and 89 fiber links

(b) Node demand: Distributed traffic model based on population statistics – DWG

(c) Node demand: Centralized traffic model based on measurements – DFN

Fig. 2. Network and demands

count in 14 subsequent months. For every source-destination pair we took the maximum of the 14 corresponding values. In the resulting matrix Frankfurt is the dominating traffic source [4, 9], see Fig. 2(c).

- The decentralized DWG matrix is calculated based on a geographical population model introduced by Dwivedi & Wagner in [10]. This model takes as input for every considered location – the total population, the number of (non-production) employees and the number of households (with Internet connection) – and calculates a forecast matrix based on the three service classes: voice traffic, (transaction) data traffic and Internet traffic. We used data from the German federal agency for statistics to parametrize the Dwivedi-Wagner model [11]. With such parametrization, it turns out that the DWG demands are relatively evenly distributed [4, 9], see Fig. 2(b).

**Input design parameter - traffic volume**   Both DFN and DWG matrices are scaled in such a way that the sum of the demand values for all location pairs (the total traffic volume) amounts to either 3 Tbps or 6 Tbps. We refer to the resulting four different matrices as DFN-3T, DFN-6T, DWG-3T, and DWG-6T in the following. All traffic values are converted to Gbps and rounded up if fractional. While by definition the sum of all demands is identical in DFN-3T and DWG-3T, the sum of demands terminated at a single node ranges between 13 Gbps and 1526 Gbps in DFN-3T compared to 69 Gbps and 541 Gbps in DWG-3T.

**Input design parameter - circuit speed**   The circuit speed is the rounded capacity (in Gbps) of the linecard ports in the IP and the WDM layer, respectively. For simplicity, we assume that matching circuit speeds are always available in IP and WDM layers. Within this study we consider two different circuit speeds: 10 Gbps and 100 Gbps. We chose to avoid the 40 Gbps denomination step because we estimated that vendors chose to stagger 40 Gigabit Ethernet products beyond the first wave of 100 Gigabit Ethernet shipments and therefore many networks will be migrating directly from 10 Gbps to 100 Gbps lambdas.

**Input design parameter - WDM circuit cost**   Equipment cost clearly has an influence on the total capital cost of a network. To investigate the possible link between relative WDM versus IP layer cost ratio and optimal network transparency, we chose to multiply the cost of WDM circuits by scaling transponder costs (2, 5, 10, 20 and so on).

### 3. Computational Study

We used the network architectures, cost, capacity, and traffic models described in *Section 2* to parametrize integer programming models as in [4, 9]. Models and solution methods are described in *Appendix B* in full detail. These models have been optimized using CPLEX 12.2 [12] for different scenarios.

Our aims were to understand how the network changes from capital as well as structural viewpoints. To this end, we chose to vary design parameters that might shift during volume, traffic affinity and technology changes, namely:

- 3T to 6T – increased traffic volume (in Tbps)
- DWG versus DFN – Shifting traffic patterns (decentralized versus centralized)
- 10G versus 100G – Link speed upgrade in WDM and IP layers (in Gbps)

In our list of results, we encode each optimized scenario with the relevant abbreviation, for example 10G-DFN-6T represents the scenario where only 10 Gbps ports are admissible and a total volume of 6 Tbps has been distributed according to the measurements in DFN (centralized traffic model).

### 3.1. Comparison of Node Architectures

In *Section 2* we outlined two basic node architectures: (i) "transparent core" scenario where IP-core routers are *not* present and traffic between IP-edge routers is routed 100% transparently over optical channels (see Figure 1(a)), and (ii) – the more common architecture where IP core routers are present and thus the network can be configured opaque at least to some degree (see Figure 1(b)). From the practical standpoint, these two node architectures are mutually exclusive and decision between them must be made early. Also, we make an assumption that the technical process of offering and managing lambda circuits versus subwavelength label-switched paths (LSPs) makes it impractical to mix service endpoints of dissimilar types. In addition, we do not consider the use of muxponders in both architectures because that would require a separate low-speed optical multiplexing and optical cross-connect network.

The fully transparent architecture corresponds to a 1-layer optimization problem in the optical layer as there is no IP-layer. Moreover, it results in the cheapest architecture whenever all IP demands in the traffic matrix are identical in capacity to $N$ times the capacity of the optical circuits (where $N$ is a positive integer). In such an extreme traffic scenario all IP-routers in any optimal hybrid solution have opacity level zero – that is, there is no IP transit traffic. Grooming and traffic mixing would only yield additional costs and hence IP-routers would be redundant. However, a comparison of the different architectures against non-trivial demand matrices produces a more complex picture.

Table 1 provides a cost comparison of optimal transparent and optimal hybrid architectures with respect to different demand matrices and circuit speeds. We distinguish the cost of the core equipment as indicated by the dashed box in Figure 1(a) and Figure 1(b) and the total cost which includes the cost for edge interfaces.

From Table 1, we can see that a transparent core architecture offers significant savings in DFN-3T and DWG-3T cases when using 10Gbps optical circuits. However, this architecture misbehaves with respect to the network evolution scenarios – if using 10 Gbps circuits, it runs out of available lambdas when the traffic increases (3T to 6T case) as indicated by "not feasible" cells in Table 1. When utilizing 100 Gbps circuit speeds, a fully transparent network design becomes more expensive for a 3 Tbps demand matrix and does not offer significant cost advantages over hybrid/optimized architecture in a 6 Tbps case.

Therefore, the transparent core architecture remains challenging to recommend because it exhibits splintered behavior in common evolution scenarios – such as demand growth and circuit speed upgrades. Without intermediate subwavelength multiplexing, grooming efficiency for low-speed flows goes down and lambda proliferation goes up, potentially stretching the available wavelength spectrum.

This conclusion is in line with our knowledge of carrier network designs [13] [14], where an electrical multiplexing and transit layer (typically in the form of IP core routers) is almost invariably present. In the rest of this paper we are only considering the hybrid/optimized architectures.

| Node architecture | 10G | | | | 100G | | | |
|---|---|---|---|---|---|---|---|---|
| | DFN-3T | DWG-3T | DFN-6T | DWG-6T | DFN-3T | DWG-3T | DFN-6T | DWG-6T |
| Transparent core | 2138 | 2109 | not feasible | not feasible | 3020 | 2975 | 3410 | 3275 |
| Optimized core | 4514 | 4419 | 8745 | 8629 | 4492 | 4355 | 8011 | 8063 |
| Transparent edge | 2811 | 2760 | not feasible | not feasible | 6563 | 6428 | 8437 | 8142 |
| Optimized edge | 1900 | 1900 | 3800 | 3800 | 1900 | 1900 | 3800 | 3800 |
| Transparent total | 4949 | 4869 | not feasible | not feasible | 9583 | 9403 | 11847 | 11417 |
| Optimized total | 6414 | 6319 | 12545 | 12429 | 6392 | 6255 | 11811 | 11863 |
| Difference | +29.6% | +29.8% | | | -33.3% | -33.5% | +0.3% | +3.9% |

Table 1. Capital Network Cost for Transparent Core Versus Optimized Node Architecture

## 3.2. Cost Dynamics of Hybrid/Optimized Node Architecture

| Scenario | 10G | | | | 100G | | | |
|---|---|---|---|---|---|---|---|---|
| | DFN-3T | DWG-3T | DFN-6T | DWG-6T | DFN-3T | DWG-3T | DFN-6T | DWG-6T |
| # lambdas | 387 | 388 | 762 | 763 | 51 | 54 | 89 | 91 |
| # IP paths | 216 | 263 | 227 | 262 | 164 | 185 | 185 | 218 |

Table 2. IP and WDM Routing in Optimized Node Architecture

When looking strictly into hybrid design of our reference network (Table 1 and Table 2), we can observe the following:

- Capital costs remain almost linear with respect to traffic growth. The change in demand from 3 Tbps to 6 Tbps results in a 95% cost increase for 10 Gbps circuits and between 85 and 90% for 100 Gbps circuits. This means the currently available IP and WDM equipment can economically accommodate the traffic growth scenario that we projected.

- There is no clear influence of traffic affinity (DFN versus DWG) on network cost. This conclusion is in contrast to results presented in [4] that were based on exactly the same traffic requirements and distributions, but an earlier cost model [6]. In that former work there is a considerable cost jump from a 640 Gbps IP router to an IP router with 1280 Gbps capacity. Moreover, the optimized centralized (DFN) scenario in [4] required fewer multichassis systems than the more distributed (DWG) scenario. The IP routers considered in this study however comprise an updated single-chassis capacity threshold of 2240 Gbps, which allows more homogeneous use of IP routers without going into multishelf configurations too often, independent of the DFN versus DWG scenario. From a practical perspective, this means that service providers concerned with demand shifts should look into vendors with the highest capacity platforms – even when such systems might not be needed at the time of initial network deployment.

- There is slight improvement in capital cost related to lambda transition from 10 Gbps to 100 Gbps speed. It appears that cheaper bandwidth in a 100 Gbps transmission case can overcome the negative effect of increased grooming inefficiency. Table 2 also clearly shows the positive effect of 100Gbps circuits on the total number of used lambdas and the total number of IP paths used for load balancing.

- Noticeable savings can be achieved if the WDM fabric allows for coexistence of 10 Gbps and 100 Gbps circuits in the same wave plan without muxponders. Although such technology is not commercially available today, we exemplarily optimized the scenario DWG-6T – allowing 10 Gbps as well as 100 Gbps wavelengths on fiber (not shown in Table 1) with resulting improvement in capital cost up to 9.1% (8629 versus 7843 cost units).

## 3.3. Dynamics of Node Opacity

Table 3 summarizes structural properties for the optimal network design. The first two rows provide absolute values (in Gbps) for the total transit traffic in the IP-layer ($F_{IP}$) and the WDM-layer ($F_{WDM}$). In both cases, the demand that must be terminated at the PoP locations is not considered – see the previous definition and *Appendix B*. Also note that the transit traffic values aggregate data for all 17 PoP locations, while intermediate nodes (additional 33 WDM installations) are not considered. The third row shows the node opacity in the cost-optimized networks as defined in equation (2). The same results can be interpreted from different perspectives – the growth of traffic, the distribution of traffic and the increase in circuit speed.

| Scenario | 10G | | | | 100G | | | |
|---|---|---|---|---|---|---|---|---|
| | DFN-3T | DWG-3T | DFN-6T | DWG-6T | DFN-3T | DWG-3T | DFN-6T | DWG-6T |
| Transit $F_{IP}$ | 180 | 210 | 188 | 205 | 1060 | 1393 | 1171 | 1375 |
| Transit $F_{WDM}$ | 4039 | 4259 | 6628 | 7447 | 6740 | 7242 | 11682 | 13549 |
| Opacity $\phi$ (%) | 4.3 | 4.7 | 2.8 | 2.7 | 13.6 | 16.1 | 9.1 | 9.2 |

Table 3. Transit Traffic in IP and WDM Domains

- The observation of opacity versus traffic reveals that the absolute IP transit volume within a fixed circuit type remains almost constant with respect to demand changes. For WDM transit traffic the situation is the opposite – with traffic growth, optical switching starts taking more flows. Thus, we conclude that with fixed circuit speed, increased traffic volume results in a trend toward increased optimal node transparency. The reciprocal opacity parameter $\phi$ decreases with a shift from 3T to 6T in 10 Gbps (-34.9% DFN to -42.6% DWG) as well as in a 100 Gbps case (-33.1% DFN to -42.9%DWG).

- From Table 3 it can be also be concluded that the optimal network topology is robust against variations in traffic gravity. There are only small differences in the opacity between DFN and DWG demand structures.

- Upgrading the network from 10 Gbps to 100 Gbps circuit speed results in markedly increased optimal opacity, with up to 13 through 16% of all transit traffic passing through IP core routers. The IP transit traffic $F_{IP}$ increases by roughly a factor of 6.4 while the WDM transit traffic $F_{WDM}$ increases only by a factor of 1.4 from 10G to 100G. These numbers are independent of the total amount and distribution of traffic (3T or 6T, DFN or DWG). For many nodes the WDM transit goes even down to zero. A deeper look into the underlying node-granular optimization results – not shown in Table 3 – reveals that there is a clear trend toward "dedicated" IP aggregation nodes (with 100% opacity) in the 100 Gbps scenario. It is interesting to note that such nodes tend to gravitate towards non-central locations with relatively small local (add-drop) IP traffic, such as Norden and Ulm in the Germany network, compared with Figure 2(b) or Figure 2(c). In contrast, important central nodes such as Hannover and Frankfurt remain relatively transparent (small opacity for 10G and also 100G).

### 3.4. Optimal Node Opacity as a Function of Relative IP and WDM Equipment Cost

In the telecom industry, it is not uncommon to assert that the use of optical bypass (node transparency) should be driven by the relative cost of IP router ports in connection to that of optical cross-connect switches and WDM terminals. For example, in [15], the cost of a 10 Gbps transponder is compared to IP router port cost with the conclusion that node transparency should increase to overcome high IP router port prices.

To test this hypothesis, we have produced several computations for 10 Gbps and 100 Gbps designs, where the cost of lambda is progressively increased by scaling transponder pricing. Our expectation was to see the increasing node opacity as the optical layer becomes more expensive and the optimization process resorts to aggregating traffic in IP routing nodes for minimum use of light paths. The results are given in Table 4 which shows the network cost as well as traffic characteristics and the number of installed lambdas with respect to increasing transponder cost. The default transponder cost amounts to 1.0 in the 10 Gbps network and 8.0 in the 100 Gbps case respectively (see *Appendix A*).

| Scenario | 10G-DWG-6T | | | | | 100G-DWG-6T | | | | |
|---|---|---|---|---|---|---|---|---|---|---|
| Transponder cost | 1.0 | 2 | 5 | 10 | 20 | 8.0 | 20 | 50 | 100 | 200 |
| Cost (with edge) | 12429 | 13951 | 18545 | 26174 | 41427 | 11863 | 14258 | 19487 | 28355 | 46108 |
| IP transit $F_{IP}$ | 205 | 237 | 205 | 237 | 231 | 1375 | 1293 | 1293 | 1210 | 1159 |
| WDM transit $F_{WDM}$ | 7447 | 6914 | 6566 | 7347 | 7104 | 13549 | 14761 | 14385 | 12259 | 13914 |
| Opacity $\phi$ | 2.7 | 3.3 | 3.0 | 3.1 | 3.1 | 9.2 | 8.1 | 8.2 | 9.0 | 7.7 |
| # lambdas | 763 | 763 | 763 | 763 | 763 | 91 | 90 | 90 | 89 | 89 |

Table 4. Optimal Network Opacity as Function of Transponder Cost Scaling

Surprisingly transponder cost seems to have very small effect on the optimal degree of opacity in both 10 Gbps and 100 Gbps cases. This remarkable stability signifies the fact that the optimization algorithm in the default set of our model parameters does not exactly target the reduction of transit IP ports. Instead, it works to minimize the overall number of optical-electrical-optical (OEO) conversion points in the entire network. Therefore, the optimal amount of optical bypass in our reference network is largely independent of IP versus WDM cost debate and converges to the least OEO conversions.

### 4. Conclusions and Future Work

In this study, we considered the subject of optimal network transparency using a well-known, fiber-rich network topology. Within parameters and limitations of this reference network, we observed the following:

- Despite the freedom to use non-blocking WDM switches in any locations to serve relatively few IP/MPLS nodes (17 out of 50), we did not find the fully transparent (100% optical switching) network design to be optimal with respect to common network evolution scenarios. Presence of electrical subwavelength multiplexing (in the form of IP core routers) was needed to make sure the network could scale predictably across volume and circuit speed changes. Even when volume of transit IP traffic is low, core routers were found effective in aggregating a large number of irregular-sized flows and keeping them away from consuming too many wavelengths.

- We could not prove the hypothesis that optical bypass is driven by the lower capital cost (price per bit) of optical transponders versus IP router ports. In fact, we have observed that the optimal degree of node transparency is virtually immune to relative pricing of IP layers versus optical layers.

- We observed that the optimal network opacity is mostly defined by interplay of traffic volume and available circuit speeds. Moreover, we found that under the changing conditions, some nodes remain predominantly opaque, suggesting there might be good reasons for placing higher-capacity routers at well-connected locations with relatively little local demand.

- In our reference network, we found a small but robust incentive to use 100 Gbps circuits in place of 10 Gbps links. We also confirmed the fact that 100 Gbps transmission effectively conserves lambdas and reduces the need for IP multipath load balancing.

The results of our study suggest that future research on network modeling and optimization should primarily focus on traffic and circuit characteristics. In particular, we need to check how the optimized network may change after departure from a static, constant bit-rate traffic model. Another interesting area of study is the impact of variable optical channel bandwidth allocation technology (also known as "Flexible Grid" [16]) that enables an optical network to support an arbitrary mix of wavelength speeds without capacity degradation.

**References**

1. J. Simmons and A. Saleh, "The value of optical bypass in reducing router size in gigabit networks," in "Communications, 1999. ICC'99. 1999 IEEE International Conference on Communications,", vol. 1 (IEEE, 1999), vol. 1, pp. 591–596.
2. T. Engel, A. Autenrieth, and J.-C. Bischoff, "Packet layer topologies of cost optimized transport networks Multi-layer network optimization ," in "Proceedings of the ONDM," (2009), pp. 1–7.
3. "Report on Resilience Mechanism for NOBEL Solutions in Medium and Long Term Scenarios," http://www.ist-nobel.org/Nobel2/imatges/D2[1].3_v1.1am.pdf.
4. A. Bley, U. Menne, R. Klähne, C. Raack, and R. Wessäly, "Multi-layer network design – A model-based optimization approach," in "Proceedings of the 5th PGTS 2008," (Berlin, Germany, 2008), pp. 107–116.
5. "IST Project NOBEL – Phase 2," http://www.ist-nobel.org/.
6. R. Hülsermann, M. Gunkel, C. Meusburger, and D. Schupke, "Cost modeling and evaluation of capital expenditures in optical multilayer networks," Journal of Optical Networking **7**, 814–833 (2008).
7. "DFN-Verein, German National Research and Education Network," http://www.dfn.de.
8. S. Orlowski, M. Pióro, A. Tomaszewski, and R. Wessäly, "SNDlib 1.0–Survivable Network Design Library," in "Proceedings of the 3rd International Network Optimization Conference (INOC 2007), Spa, Belgium," (2007). http://sndlib.zib.de.
9. F. Idzikowski, S. Orlowski, C. Raack, H. Woesner, and A. Wolisz, "Dynamic routing at different layers in IP-over-WDM networks – Maximizing energy savings," Optical Switching and Networking, Special Issue on Green Communications **8**, 181–200 (2011).
10. A. Dwivedi and R. E. Wagner, "Traffic model for USA long distance optical network," in "Proceedings of the Optical Fiber Communication Conference," (2000), pp. 156–158.
11. "Statistisches Bundesamt Deutschland," http://www.destatis.de/jetspeed/portal/cms/.
12. IBM, "IBM ILOG CPLEX Optimizer," .
13. G. Li, D. Wang, R. Doverspike, C. Kalmanek, and J. Yates, "Economic Analysis of IP/Optical Network Architectures," Optical Fiber Communication Conference (2004).
14. S. Elby, "Core Network Transformation," (2007).
15. G. Bennett, "Next-Generation IP over WDM Network Architectures," .

16. S. Gringeri, B. Basch, V. Shukla, R. Egorov, and T. J. Xia, "Flexible architectures for optical transport nodes and networks," IEEE Communications Magazine **48**, 40–50 (2010).
17. S. Raghavan and D. Stanojević, "WDM Optical Design using Branch-and-Price," (2007). Working paper, Robert H. Smith School of Business, University of Maryland.
18. R. Ahuja, T. Magnanti, and J. Orlin, *Network Flows: Theory, Algorithms, and Applications* (Prentice Hall, 1993).
19. A. Koster, S. Orlowski, C. Raack, G. Baier, and T. Engel, *Single-layer Cuts for Multi-layer Network Design Problems* (Springer-Verlag, College Park, MD, U.S.A., 2008), vol. 44, chap. 1, pp. 1–23. Selected proceedings of the 9th INFORMS Telecommunications Conference.
20. B. Fortz and M. Thorup, "Internet traffic engineering by optimizing OSPF weights," in "Proceedings of the INFOCOM, Tel-Aviv, Israel," (2000).

### A. Device Definitions and Costs

We consider a WDM layer to consist of optical cross-connects (OXCs) devices that are interconnected by links representing optical fibers. Each fiber can carry up to 40 WDM channels. Devices in the IP layer correspond to IP-routers that are interconnected by light paths (lambdas) provided by the WDM layer. A light path has a capacity of either 10Gbps or 100Gbps. A light-path can be established using arbitrary paths in the WDM network, and we do not consider penalties related to color direction conflicts or capacity reservation for resiliency (path restoration is left to the upper layer). We assume that if traffic passes through an IP/MPLS node it can be arbitrarily split, multiplexed, or routed over parallel links with normal IP/MPLS path computation and load-sharing algorithms.

For optical cross-connect equipment, we distinguish optical cross-connects with different fiber degrees that determine the maximum number of fibers that can be connected in the WDM layer. An optical cross-connect might couple incoming WDM channels to outgoing ones (performing optical bypass), multiplex channels, or terminate them toward the IP layer, which essentially means that optical signals can be arbitrarily switched. According to terminology introduced in [6] we call the corresponding multiplexer/demultiplexer port together with optical amplifiers a WDM-terminal. A WDM-terminal is installed at the optical cross-connect for every connected fiber.

At both ends of a light-path we assume a transponder to be installed at the add-drop port of the terminating optical cross-connect. This transponder is responsible for transmitting the colored optical signal carrying 10 Gbps or 100 Gbps capacity. In our studies we assumed long-haul (LH) transponders with a maximum reach of 750 kilometers – that is, a light path cannot be longer than 750 kilometers without full regeneration, which is determined by the total length of the used fibers on the path in the WDM network. Every transponder is connected to a gray short-reach (SR) interface on the IP router.

Apart from transponders there is some additional equipment we need to add to properly establish and recondition optical signals over large distances – namely optical line amplifiers (OLA). We assume that amplifiers are already installed at every optical cross-connect. In addition dynamic gain equalizers (DGEs) are needed at every fourth amplification site. To compensate for effects caused by chromatic dispersion, the so-called dispersion compensating fibers (DCFs) are provided at amplification sites. The amount of DCF needed depends on the length of the link.

The cost model introduced in [6] provides prices for IP and WDM equipment normalized to the cost of a 10 Gbps LH transponder (which is defined at cost 1.0). We leave this model and corresponding price relations unchanged for the WDM equipment. The optical hardware used in our study including all cost values is summarized in Table 5. We also add an LH transponder with capacity 100 Gbps at cost 8.0. This reflects the fact that currently a 100Gbps transponder costs approximately the same as eight separate 10 Gbps units.

For every IP node, we adopt equipment from two different vendors: one full-rack size router with 16 slots, up to 140 Gbps (14x10G interfaces or 1x100G interface) per slot and multichassis option; another router in one-third rack format with 11 120 Gbps slots (supporting 12x10G or 1x100G interfaces) – but without the availability of a multichassis extension shelf. For 100 Gigabit Ethernet we estimated linecards to be available approximately at the same cost as six 10 Gigabit Ethernet (WAN Phy) ports and 100G SR transceivers at four times the cost for 10G SR transceivers. We based our estimates on original equipment quotes with delivery dates in 2012. To establish a relation between IP and WDM equipment cost, we conducted a survey of equipment vendors and network operators, and we found that in the last five years the cost per Gbps has decreased more significantly for IP equipment compared to WDM equipment – which is reflected in Table 6. The larger router (Type1) at base cost of 27.25 has full-duplex switching capacity of 2240 Gbps that corresponds to the lowest price of 1.9 cost units per 10G port (when fully configured). The smaller router (Type2) at base cost 12 has full-duplex switching capacity of 1320 Gbps and 10G cost of 1.67 units

| Device | Capacity, ports | Cost | Remarks |
|---|---|---|---|
| transponder 10G LH | 10Gbps | 1.0 | |
| transponder 100G LH | 100Gbps | 8.0 | original estimate |
| ROADM 50% | fibers: 2, channels add-drop: 40 | 11.67 | incl. 2 WDM terminals |
| ROADM 100% | fibers: 2, channels add-drop: 80 | 17.5 | incl. 2 WDM terminals |
| OXC 3–5 | fibers: $n \in \{3,\ldots,5\}$, add-drop: $40 \cdot n$ | $2.5 + n \cdot 8.33$ | incl. $n$ WDM terminals |
| OXC 6–10 | fibers: $n \in \{6,\ldots,10\}$, add-drop: $40 \cdot n$ | $2.75 + n \cdot 8.99$ | incl. $n$ WDM terminals |
| WDM terminal | fibers: 1 | 4.17 | including mux/demux, OLA |
| DGE | fibers: 1 | 2.17 | every 320 km fiber span |
| DCF | fibers: 1 | 0.0072 | per km fiber spam |
| OLA | fibers: 1 | 1.92 | every 80 km fiber span |
| Fiber | channels: 40 | – | indirect cost (DGE, DCF, OLA) |

Table 5. DWDM Cost Model (40 Channel System)

(partial router configurations yield higher effective port cost).

We ignore the base cost for edge routers throughout this paper as it is not dependent on the network design. However, we consider the core-facing interfaces on edge routers (see Figure 1(a) and Figure 1(b)) as well as edge-facing interfaces on IP-routers (see Figure 1(b) This is necessary to reflect the way the two different architectures affect the edge connectivity. For example, in the optimized architecture we can choose 10 Gbps interconnects between core and edge routers even if the former are interfacing only 100G circuits. This would be impossible in the "fully transparent" core scenario. Whenever edge interfaces are included, we consider them to be of the same type as linecards from router Type2.

| Device | Capacity, slots, ports | Cost |
|---|---|---|
| IP-router Type1 2240G incl. software | 2240Gbps, slots: 16 | 27.3 |
| Multichassis (fixed-charge cost) | Connects Type1 routers | 50.0 |
| slot-card 140G | 140Gbps | 14.0 |
| line-card 14x10G | ports: 14 at 10Gbps | 11.0 |
| line-card 8x10G | ports: 8 at 10Gbps | 8.0 |
| line-card 1x100G | ports: 1 at 100Gbps | 6.0 |
| IP-router Type2 1320G incl. software | 1320Gbps, slots: 11 | 12.0 |
| slot-card/line-card 12x10G | ports: 12 at 10Gbps | 19.0 |
| slot-card/line-card 1x100G | ports: 1 at 100Gbps | 14.0 |
| SR transceiver 10G | 1 port | 0.5 |
| SR transceiver 100G | 1 port | 2.0 |

Table 6. IP Router Cost Model

## B. Mathematical Model

Hereby we aim at describing the model and solution approach used to solve the cost minimization problem arising from the considered network architecture together with the hardware and cost model introduced in the previous section. In particular we highlight the chosen level of abstraction and explain to which extent we implemented the relevant practical side constraints.

The mixed-integer programming model used in our optimizations is an extension of the formulation proposed for instance in [4,9,17]. It has the advantage of a very compact description of the flow on the virtual layer. This is achieved by aggregating all virtual flow variables corresponding to the same node pair and all demands with the same source node to a single variable.

We model the potential physical (optical dark fiber) topology depicted in Figure 2(a) as an undirected graph $G = (W, E)$ with nodes $W$ and edges $E$, that is in our case $|W| = 50$ and $|E| = 89$. The potential virtual (IP) topology is given by $H = (V, V \times V)$, where $V$ denotes a subset of $W$ with $|V| = 17$ and $V \times V$ is the set of all corresponding node pairs with $|V \times V| = 136$, that is, the virtual layer is the complete undirected graph spanned by the 17 IP demand locations. A *virtual link* is just a pair of nodes and has potentially different realizations in the physical domain.

We are given a set $D \subseteq V \times V$ of demands (or commodities) between the 17 IP demand locations. Each demand $k \in D$ has a source $s(k) \in V$, a target $s(k) \in V$ and a value $d^k$. These values correspond to the DFN and DWG demand matrices and have to be realized in the virtual IP layer $H$. This is modeled by introducing a so-called multi-commodity flow [18]. For every node pair $(i, j)$ the continuous flow-variables $f_{ij}^k \in \mathbb{R}_+$ and $f_{ji}^k \in \mathbb{R}_+$ describe the flow between $i$ and $j$ in both directions corresponding to demand $k \in D$. For every demand $k \in D$ the flow conservation constraints

$$\sum_{j \in V \setminus \{i\}} (f_{ij}^k - f_{ji}^k) = \begin{cases} d^k & i = s(k) \\ -d^k & i = t(k) \\ 0 & \text{else} \end{cases}, \quad \text{for all } i \in V, \qquad (3)$$

ensure that every demand realizes between source $s(k)$ and target $t(k)$ and that no flow disappears or enters the network from outside. The number of flow-variables and constraints (3) can be reduced considerably by aggregating demands that have the same source node (or the same target node) – see for instance [9, 19]. This aggregation is done for our computations. Also notice that in our IP-flow model the flow is aggregated between virtual nodes rather than being described on individual light paths. Mapping the virtual flow to light paths can be done easily in a post-processing step. As there is no additional assumption on the IP flow we essentially assume that IP traffic can be arbitrarily split and routed via multiple virtual paths that could be enabled for instance by MPLS (multi-protocol label switching). We remark, however, that the main motivation behind this assumption is computational tractability since any more restrictive routing assumption (for example, single-path flow or path length restrictions) would involve using discrete flow variables and/or additional flow constraints. From the practical point of view, we refer the reader to [20] for a discussion on the similarity of load distribution in a network using fractional multi-commodity flows and a network using the OSPF protocol with ECMP (equal-cost multipath) routing.

For every virtual link $(i, j) \in V \times V$ a set $P_{(i,j)}$ of admissible paths in the physical network is considered. As we assume long-haul (LH) equipment, the length of a path (the sum of the edge lengths in km) is restricted to 750 km. In case that for some node pair there were more than 50 long haul paths available, we used the 50 shortest paths in our calculations – that is, $|P_{(i,j)}| \leq 50$ for every node pair $(i, j) \in V \times V$. Let $P$ be the union of all these paths. In our case $|P| = 5591$. Each path $p \in P$ can be equipped with capacity modules having of bit-rate $A$ which we refer to as being the routing capacity. Such a capacity module corresponds to a light path in the optical network and consists of the two transponders at the end-nodes of the light path used to establish the optical signal and the two consumed ports (SR transceivers) at the IP router line-card. The integer variable $y_p \in \mathbb{Z}_+$ denotes the number of light paths established on path $p \in P$ such that the total capacity of a path $p$ is given by $Ay_p$ and the total capacity of a virtual link $(i, j) \in V \times V$ by $\sum_{p \in P_{(i,j)}} Ay_p$ which corresponds to the capacity of all light paths between $i$ and $j$. The total virtual link flow (for all demands) cannot exceed the provided virtual link capacity that is reflected by the following virtual link capacity constraints:

$$\sum_{k \in D} (f_{ij}^k + f_{ji}^k) \leq \sum_{p \in P_{(i,j)}} Ay_p, \quad \text{for all } (i, j) \in V \times V. \qquad (4)$$

Every light path installed on the physical path $p \in P$ consumes one wavelength channel on every edge $e \in p$. Wavelength channels are provided by fibers to be activated on physical links. A physical link can be equipped with an arbitrary number of fibers each supporting a total of $B$ channels. We denote by $y_e \in \mathbb{Z}_+$ the variable counting the number of fibers for every physical link $e \in E$. Physical link capacity constraints ensure that the channel capacity is not exceeded:

$$\sum_{p \in P : e \in p} y_p \leq By_e, \quad \text{for all } e \in E \qquad (5)$$

For nodes we introduce the sets $N$ and $O$ of admissible virtual and physical node modules, respectively. Only one virtual node module and one physical node module can be installed at every location while virtual node modules are restricted to the locations $V$. These constraints are realized by imposing the following:

$$\sum_{n \in N} x_i^n \leq 1, \quad \text{for all } i \in V, \quad \text{and} \quad \sum_{o \in O} x_i^o \leq 1, \quad \text{for all } i \in W, \qquad (6)$$

The the binary variable $x_i^n \in \{0, 1\}$ determines whether module $n$ is installed at node $i \in V$ or not. Similarly, the binary variable $x_i^o \in \{0, 1\}$ determines whether or not to install the physical node module $o \in O$ at node $i \in W$. A virtual node module $n \in N$ corresponds to a preconfigured IP router. It has a maximum switching capacity $C^n$ in Gbps and a maximum number of available slots $S^n \in \mathbb{Z}_+$. Every light path corresponding to a path $p \in P$ consumes $A'$ Gbps of switching capacity and a portion $s \in \mathbb{R}_+$ of slots at the two virtual node modules installed at the end-nodes of $p$. Notice that for a light path the switching-capacity $A'$ consumed at the node module can be different from the routing capacity

$A'$ provided to the virtual IP flow – see the following – but it holds that $A' \geq A$. For a node $i \in V$ let the set $\delta_P(i) \subset P$ denote all physical paths having one of its end-nodes in $i$. The virtual node capacity constraint

$$d(i) + \sum_{p \in \delta_P(i)} A' y_p \leq \sum_{n \in N} C^n x_i^n, \quad \text{for all } i \in V \qquad (7)$$

guarantees that node $i$ has enough virtual capacity to terminate the node demand $d(i) := \sum_{k \in D, i=s(k)} d^k + \sum_{k \in D, i=t(k)} d^k$ and to switch all traffic corresponding to paths ending at $i$. Similarly, the virtual node slot constraint

$$\sum_{p \in \delta_P(i)} s y_p \leq \sum_{n \in N} S^n x_i^n, \quad \text{for all } i \in V \qquad (8)$$

ensures that there are enough slots available at the IP-router for node $i$.

A physical node module $o \in O$ corresponds to a preconfigured optical cross connect and is determined by the number of supported fibers $F^o$ and the maximum number of add-drop ports $P^o$. Every terminated light path consumes one add-drop port. Consequently, there are two physical node capacity constraints. The fiber constraint:

$$\sum_{e \in \delta_E(i)} y_e \leq \sum_{o \in O} F^o x_i^o, \quad \text{for all } i \in W, \qquad (9)$$

which has $\delta_E(i)$ denoting all physical links ending in $i$. The add-drop channel constraint is as follows:

$$\sum_{p \in \delta_P(i)} y_p \leq \sum_{o \in O} P^o x_i^o \quad \text{for all } i \in W. \qquad (10)$$

The total network installation cost is given by the cost for all virtual and physical nodes as well as link modules. The cost for virtual link modules is independent of the actual virtual link. Similarly the cost for physical and virtual node modules is independent of the node at which they are installed. However, the cost for physical link modules (corresponding to fibers) depends on the length of the corresponding physical link. We denote by $\alpha$ the cost of a virtual link module. A physical link module installed at edge $e \in E$ is given by $\beta_e$. The cost for virtual node module $n \in N$ is $\gamma^n$ and the cost for physical node module $o \in O$ is denoted by $\delta^o$ such that the total network cost amounts to:

$$\sum_{p \in P} \alpha y_p + \sum_{e \in E} \beta_e y_e + \sum_{i \in V} \sum_{n \in N} \gamma^n y_e + \sum_{i \in W} \sum_{o \in O} \delta^o x_i^o. \qquad (11)$$

The value (11) is minimized over all possible network configurations – that is, all solutions satisfying constraints (3)-(10) plus the mentioned integrality and bound constraints for the variables $y_p, y_e, x_i^n, x_i^o$.

**Transparent core solution** To calculate the optimal transparent core solution based on nodes of type Figure 1(a) we used a variation of the model presented previously. We first fixed the cost for IP routers to zero, that is, $\gamma^n = 0$ for all $n \in N$. Instead of 50 available paths per virtual link $(i, j) \in V \times V$ we only generated the shortest path between $i$ and $j$ in the optical network, that is, we set $|P_{(i,j)}| = 1$. To ensure that no traffic is switched in the virtual domain we made sure that all flow is sent on a single virtual hop by fixing $f_{ij}^k = d^k$ if $i = s(k)$ and $j = t(k)$ and $f_{ij}^k = 0$ in all other cases.

**Edge cost** We are not considering the base cost of edge routers, as it is not dependent on the optimal core topology. However, edge routers might need different interfaces towards the core and a different number of interfaces in architecture Figure 1(a) versus architecture Figure 1(b). By the aforementioned construction, the core-facing linecards are already included in the cost of the transparent core solutions. For the optimized/hybrid case we can include the additional costs (core-facing and edge-facing linecards) safely after the core optimization stage. The amount of needed interfaces only depends on the demand at the considered location.

**Cost and capacity parametrization** We have introduced the concept of link and node modules both for the virtual IP layer and the physical DWDM layer. Every instance of these entities combines a set of devices from Tables 5 and 6. Here we provide a detailed parametrization of the cost model given the constraints (3)-(10). For simplicity, all demands, capacities and flows in our model are further listed in Gbps.

Virtual node modules are preconfigured IP-routers. A virtual node module comprises the IP-router itself plus a certain number of preconfigured slots, which refers to installing the same number of slot cards and line cards (without

SR transceivers). We allow for single or multichassis router configurations with a total full-duplex switching capacity of 8960 Gbps in card 64 slots (only the larger IP router Type1 allows for such capacity). We do not consider routers Type1 for configurations between 1 and 10 slots because such densities are better served with router Type2. In total this gives us 66 different virtual node configurations installable at every IP node $i \in V$. The switching capacity of a virtual node module is determined by the number of preconfigured slots with either 120 Gbps (router Type2) or 140Gbps (router Type1) cards. The cost of a preconfigured slot is set to 16 for the small router (slot and line-card combination) and 22 for the big router. (Notice that we ignore 3 cost units per slot in the 10 Gbps case. This cost is added to the configuration in a post-processing step. In the 100 Gbps case a preconfigured slot already accounts for the necessary 100G SR transceiver.) As an example, the cost of a node module $n \in N$ with capacity $C^n = 4900$ Gbps and $S^n = 35$ slots is $\gamma^n = 901.75$. It is based on a multichassis configuration using three of the bigger IP routers and using 35 out of 48 slots – that is, we have three routers at cost 27.5 each, one multi-chassis configuration at cost 50 and 35 preconfigured slots at cost 22.

Virtual link modules correspond to light paths. A light path in the 10G case is set up using two transponders and two transceivers at the end-nodes of the path. In the 100G case it consists of two transponders. The cost of a 10Gbps respectively 100Gbps lambda thus amounts to $\gamma^n = 3$ respectively $\gamma^n = 16$. The switching or routing capacity of a 10 Gbps light path is set to $A = A' = 10$. Such a light path configuration consumes $s = 1/14$ slots at the IP-router on both sides (independent of the router type). The switching capacity of a 100Gbps light path is set to $A' = 120$ while its routing capacity is only $A = 100$. Notice that a 100G line-card blocks a complete slot at the IP router. Accordingly, we set $s = 1$ for 100G lambdas.

Regarding physical node modules there is a one-to-one correspondence to the definition of optical cross-connects and ROADMs from the NOBEL cost-model [5, 6] that we have summarized in Table 5. All nodes $i \in W$ can be equipped with one node module $o \in O$ corresponding to one of the optical cross-connects or one of the ROADMs from Table 5. The corresponding number of allowed fibers $F^o$ and add-drop ports $P^o$ as well as the cost $\delta^o$ of the physical node module can also be found in the table. Also following the NOBEL cost-model, in order to operate a fiber on a physical link $e \in E$ with length $l_e$ in km, we have to provide $N_{ola} := \lceil l_e/80 \rceil - 1$ many optical line amplifiers and $N_{dge} := \lceil N_{ola}/4 \rceil - 1$ many digital gain equalizers. In addition there is a cost of 0.0072 per km for dispersion-compensating fiber. These costs add up to the length dependent fiber cost $\beta_e$.

**Network traffic and node measures**  In *Section 2.2* the notion of node opacity has been introduced. Given the notation introduced in this section we can now formalize the input values of the definition, that is, $F_{IP}(i)$ and $F_{WDM}(i)$.

Recall that the total demand that is terminated at node $i$, or similarly the total flow leaving/entering the network at node $i \in V$, is given by the following:

$$d(i) := \sum_{k \in D, i=s(k)} d^k + \sum_{k \in D, i=t(k)} d^k.$$

Let $f_p$ be the total flow on path $p \in P$ in both directions with respect to all commodities. This value is not explicitly available in model (3)-(10) since flow is defined on pairs of nodes and not on individual paths. However, as already mentioned, given a solution, we can always disaggregate the flow from node pairs $(i, j) \in V \times V$ to the corresponding light paths $p \in P_{(i,j)}$ – that is, given a solution to model (3)-(10), we can assign a flow value to every light path without exceeding its capacity and this can be done without flow left over, which is ensured by inequality (4).

To get the total flow that is switched at the IP-router of node $i \in V$ without being terminated we have to subtract the demand flow $d(i)$ from the total terminated flow and divide by two to not count packets twice that are leaving and entering the IP-router. Accordingly, we get the following:

$$F_{IP}(i) := \left( \sum_{p \in \delta_P(i)} f_p - d(i) \right) /2$$

as the IP transit flow at node $i \in V$, also see *Section 2.2*. Remember that $\delta_P(i)$ corresponds to all physical paths that have one end node in $i$. Similarly we can evaluate all flow that optically bypasses node $i$ – that is, all flow on light paths that is switched by the cross connect at $i$ but not terminated. The corresponding WDM transit flow of node $i \in V$ is given by the following:

$$F_{WDM}(i) := \sum_{p \in P: i \in p} f_p - \sum_{\delta_P(i)} f_p.$$

We denote by $F_{IP}$ (respectively $F_{WDM}$) the total network IP (respectively WDM) transit flow where

$$F_{IP} := \sum_{i \in V} F_{IP}(i) \quad \text{and} \quad F_{WDM} := \sum_{i \in V} F_{WDM}(i).$$

Recall that node opacity is only defined for PoP locations. Together with (1) and (2) we have a meaningful definition of the network (node) opacity, see *Section 2.2*.